\documentclass[a4paper,english,aps,prb]{revtex4}
\usepackage[T1]{fontenc}
\usepackage[latin9]{inputenc}
\usepackage{units}
\usepackage{amsmath}
\usepackage{graphicx}
\usepackage{amssymb}

\makeatletter
\newcommand{\lyxdot}{.}

\@ifundefined{textcolor}{}
{%
 \definecolor{BLACK}{gray}{0}
 \definecolor{WHITE}{gray}{1}
 \definecolor{RED}{rgb}{1,0,0}
 \definecolor{GREEN}{rgb}{0,1,0}
 \definecolor{BLUE}{rgb}{0,0,1}
 \definecolor{CYAN}{cmyk}{1,0,0,0}
 \definecolor{MAGENTA}{cmyk}{0,1,0,0}
 \definecolor{YELLOW}{cmyk}{0,0,1,0}
 }
\newenvironment{lyxlist}[1]
{\begin{list}{}
{\settowidth{\labelwidth}{#1}
 \setlength{\leftmargin}{\labelwidth}
 \addtolength{\leftmargin}{\labelsep}
 }}
{\end{list}}

\makeatother

\usepackage{babel}

\begin{document}

\title{Viscoelastic Properties of Crystals}

\author{Stephen R. Williams}

\email{swilliams@rsc.anu.edu.au}

\affiliation{Research School of Chemistry, The Australian National University,
Canberra, ACT 0200, Australia}

\author{Denis J. Evans}

\affiliation{Research School of Chemistry, The Australian National University,
Canberra, ACT 0200, Australia}

\date{\today}
\begin{abstract}
We examine the question of whether fluids and crystals are differentiated
on the basis of their zero frequency shear moduli or their limiting
zero frequency shear viscosity. We show that while fluids, in contrast
with crystals, do have a zero value for their shear modulus, in contradiction
to a widespread presumption, a crystal does not have an infinite or
exceedingly large value for its limiting zero frequency shear viscosity.
In fact, while the limiting shear viscosity of a crystal is much larger
than that of the liquid from which it is formed, its viscosity is
much less than that of the corresponding glass that may form assuming
the liquid is a good enough glass former.
\end{abstract}
\maketitle

\section{Introduction}

Response theory and Green-Kubo relations provide a good understanding
of the microscopic origins of viscoelasticity in fluids \citep{Zwanzig-ARPC-1965,Kubo-ROPIP-1966,Marconi-PRSOPL-2008}.
They show that all fluids are in fact viscoelastic. However the range
of frequencies over which one sees a crossover from low frequency
viscous behaviour to high frequency elastic behaviour, varies by more
than 10 decades for various common fluids. Solids are also expected
to be viscoelastic, exhibiting viscous as well as elastic behaviour
\citep{Lakes-1998}. Of course there is a limit to the maximum strain
amplitude that can be applied to a crystal before it fractures, cleaves, plastically deforms
or melts \citep{Stankovic-PRE-2004}. However provided this limit
is obeyed one can in principle compute or measure elastic constants
at all frequencies including zero and shear viscosities at all frequencies
except zero. This presents an interesting question: how does the limiting
zero frequency shear viscosity of a crystal compare to that of a fluid
or a glass?

In contrast to fluids the microscopic origins of the rheological properties
of crystals are poorly understood. There is no equivalent to Green-Kubo
theory that has been proposed for the viscoelastic behaviour of crystals.
We know of no collection of the frequency dependent shear viscosity
of crystalline materials. There are collections of experimental data
for the frequency dependent shear moduli \citep{Lakes-1998}.

It is easy to understand why this surprising state of affairs exists.
The standard derivation of Green-Kubo expressions for the Navier-Stokes
transport coefficients relies on the Onsager regression hypothesis
and a solution of the fluctuating Navier-Stokes hydrodynamic equations
\citep{Zwanzig-ARPC-1965,evans-and-morriss-book}. For crystals the
corresponding elasto-hydrodynamic mode-coupling equations are quite
complex (and anisotropic). As far as we are aware no one has derived
the Green-Kubo relations for the frequency dependent elasto-hydrodynamic
coefficients of a crystal. Furthermore when you shear a crystal the
underlying equilibrium state varies with the strain. This is not so
in a fluid. This difference implies that the Green-Kubo derivations
themselves are inherently more complex for crystals than for the corresponding
liquid or gas.

We adopt a much easier approach to solve this problem. We restrict
ourselves to the case where there is no significant linear creep and
then employ the SLLOD equations for time dependent planar Couette
flow \citep{evans-and-morriss-book}. These equations give an exact
description of adiabatic time dependent planar Couette flow arbitrarily
far from equilibrium. These equations convert a technically complex
\emph{thermal} transport process into a much simpler mechanical process
that can be analysed using a thermostatted version of Kubo's response
theory \citep{Kubo-ROPIP-1966}.

\section{Theory}

\subsection{Thermostatted SLLOD equations for isothermal planar shear}

We use the standard isokinetic equations of motion which feature a
synthetic thermostat under the condition that the total peculiar momentum
is always zero. We know some important facts about synthetic thermostats
of this type: 
\begin{lyxlist}{00.00.0000}
\item [{1)}] Their equilibrium distribution function is known\citep{evans-and-morriss-book,Evans-arXiv-2009}. 
\item [{2)}] Any artifacts in the dynamical correlation functions due to
the synthetic thermostat are at most of order ${\cal O}(1/N)$ where
$N$ is the number of particles in the system \citep{evans-and-morriss-book}. 
\item [{3)}] The linear response of the system to an external field is
devoid of artefacts due to the synthetic thermostat \citep{evans-and-morriss-book}. 
\item [{4)}] If we wish to study nonequilibrium phenomena outside the linear
response regime we can arrange things such that the thermostat only
acts on a region far removed from the system of interest \citep{Williams-PRE-04}. 
\end{lyxlist}
The isokinetic equations of motion are\begin{eqnarray}
\dot{\mathbf{q}}_{i} & = & \mathbf{p}_{i}/m_{i}+\mathbf{\underline{C}}_{i}(\mathbf{\Gamma})\cdot\mathbf{F}_{e}\nonumber \\
\dot{\mathbf{p}}_{i} & = & \mathbf{F}_{i}(\mathbf{q})+\mathbf{\underline{D}}_{i}(\mathbf{\Gamma})\cdot\mathbf{F}_{e}-\alpha\mathbf{p}_{i}\nonumber \\
\alpha & = & \frac{\sum_{i=1}^{N}\mathbf{p}_{i}\cdot\mathbf{F}_{i}+\mathbf{p}_{i}\cdot\mathbf{\underline{D}}_{i}(\mathbf{\Gamma})\cdot\mathbf{F}_{e}}{\sum_{i=1}^{N}\mathbf{p}_{i}\cdot\mathbf{p}_{i}}\label{EOM}\end{eqnarray}
where $\mathbf{q}_{i}$ and $\mathbf{p}_{i}$ are the position and
peculiar momentum of the $ith$ particle, $\mathbf{q}$ is the $3N$
dimensional vector of all the positions, $\mathbf{\Gamma}=(\mathbf{q},\mathbf{p})$
is the $6N$ dimensional phase space vector, $m_{i}$ is the mass
of the $i^{th}$ particle, $\mathbf{F}_{i}$ is the force on the $i^{th}$
particle due to interactions with other particles, $\mathbf{F}_{e}$
is the external field which drives the system away from equilibrium.
When it is set to zero, if the system is ergodic and has decaying
memory, any arbitrary initial distribution will eventually relax to
equilibrium \citep{Evans-arXiv-2009}. $\mathbf{\underline{C}}_{i}$
and $\mathbf{\underline{D}}_{i}$ are second rank tensors which couple
the system to the external field and $\alpha$ is the thermostat multiplier
which holds the value of the peculiar kinetic energy, $\sum_{i=1}^{N}p_{i}^{2}/2m$,
constant. The equilibrium distribution function (when $\mathbf{F}_{e}=0$)
for the system is given by\begin{eqnarray}
f(\mathbf{\Gamma}) & = & \frac{\delta\boldsymbol{(}K(\mathbf{\Gamma})-K_{0}\boldsymbol{)}\delta(\mathbf{p}_{M})\exp[-\beta H_{0}(\mathbf{\Gamma})]}{Z}\nonumber \\
Z & \equiv & \int d\mathbf{\Gamma}\delta\boldsymbol{(}K(\mathbf{\Gamma})-K_{0}\boldsymbol{)}\delta(\mathbf{p}_{M})\exp[-\beta H_{0}(\mathbf{\Gamma})]\label{Eq-dist}\end{eqnarray}
where $K(\mathbf{\Gamma})$ is the peculiar kinetic energy which is
fixed to the value $K_{0}$, $H_{0}(\mathbf{\Gamma})=K(\mathbf{p})+\Phi(\mathbf{q})$
is the Hamiltonian with $\Phi(\mathbf{q})$ the potential energy due
to the particle interactions, $\mathbf{p}_{M}=\sum_{i=1}^{N}\mathbf{p}_{i}$
is the total peculiar momentum, and $\beta^{-1}\equiv k_{B}T=2K_{0}/(3N-4)$
where $k_{B}$ is Boltzmann's constant and $T$ is the equilibrium
thermodynamic temperature. We note that if the system is ergodic with
decaying memory the distribution function Eq. (\ref{Eq-dist}) is
the unique, dissipationless equilibrium state \citep{Evans-arXiv-2009}.
We also note that Eq. (\ref{Eq-dist}) includes all finite size corrections.
The Helmholtz free energy of the system is \citep{Evans-arXiv-2009}
\begin{equation}
A=-k_{B}T\,\ln[Z].\label{Helmholtz}\end{equation}

For the system undergoing planar shear we use the so called isokinetic
SLLOD equations of motion \citep{evans-and-morriss-book}, \begin{eqnarray}
\dot{\mathbf{q}}_{i} & = & \mathbf{p}_{i}/m_{i}+\mathbf{i}\dot{\gamma}(t)q_{yi}\nonumber \\
\dot{\mathbf{p}_{i}} & = & \mathbf{F}_{i}-\mathbf{i}\dot{\gamma}(t)p_{yi}-\alpha\mathbf{p}_{i}\nonumber \\
\alpha & = & \frac{\sum_{i=1}^{N}\mathbf{p}_{i}\cdot\mathbf{F}_{i}-\dot{\gamma}(t)p_{xi}p_{yi}}{\sum_{i=1}^{N}\mathbf{p}_{i}\cdot\mathbf{p}_{i}},\label{sllod}\end{eqnarray}
where $\mathbf{i}$ is the unit vector in the direction of the \emph{x}
Cartesian axis, $p_{xi}$ is the $x$ component of $\mathbf{p}_{i}$
the peculiar momentum of particle \emph{i}, and $\dot{\gamma}(t)=du_{x}/dy$
is the time dependent strain rate where $u_{x}$ is the x-component
of the steaming velocity. In computer simulations Eqs. (\ref{sllod})
are used in conjunction with Lees-Edwards shearing periodic boundaries
\citep{evans-and-morriss-book} to minimize the system size dependence
of the results. The adiabatic form of these equations give an exact
description of adiabatic shear flow arbitrarily far from equilibrium.
They are equivalent to Newton's equations of motion plus an integrated
shift in the x-laboratory velocity of $\int_{0}^{t}ds\ddot{\gamma}(s)q_{yi}(s)$
for every particle. Decomposing the strain rate into an infinite sum
of infinitesimal Heaviside steps shows that the adiabatic form of
Eq. (\ref{sllod}) is exact for time dependent planar Couette flow.

\subsection{The difference between solids and fluids under quasistatic strain}

A fundamental difference between a fluid and a solid is that while
a solid can support a small externally applied stress indefinitely,
a fluid cannot. Fluids will always flow in response to the applied
stress thereby eventually reducing the magnitude of the stress to
zero. If we subject a liquid to a quasistatic strain rate there will
be no work done in shearing it. To prove this we note that the rate
at which work is done in shearing a single ensemble member is given
by

\begin{equation}
\dot{W}(t)\equiv\dot{H}_{0}^{ad}(\mathbf{\Gamma})=\frac{\partial H_{0}}{\partial\mathbf{\Gamma}}\cdot\dot{\mathbf{\Gamma}}^{ad}\equiv-\mathbf{J}(\mathbf{\Gamma})V\mathbf{F}_{e}(t)=-\dot{\gamma}(t)VP_{xy}(\mathbf{\Gamma}),\label{work_rate}\end{equation}
where the flux, $\mathbf{J}$, introduced here is defined by Eqs.
(\ref{sllod} \& \ref{work_rate}) and $P_{xy}$ is the $xy$ element
of the pressure tensor (closely related to the shear stress, $\sigma_{xy}=-\left\langle P_{xy}\right\rangle $).
Using Eq. (\ref{sllod}) we see that\begin{equation}
VP_{xy}(\mathbf{\Gamma})=\sum_{i=1}^{N}\frac{p_{xi}p_{yi}}{m_{i}}+\sum_{i=1}^{N}F_{xi}q_{yi}.\label{Pxy}\end{equation}
To leading order we have $\left\langle P_{xy}\right\rangle \sim\dot{\gamma}+{\cal O}(\dot{\gamma}^{3})$
and so if we calculate the work required to quasistatically strain
a fluid by a fixed amount $\delta\gamma$, at constant shear rate
$\dot{\gamma}$, we obtain\begin{equation}
\left\langle \Delta W\right\rangle _{QS}=-\lim_{\dot{\gamma}\rightarrow0}\,\dot{\gamma}V\int_{0}^{\delta\gamma/\dot{\gamma}}dt\,\left\langle P_{xy}(t)\right\rangle =\lim_{\dot{\gamma}\rightarrow0}V\eta\dot{\gamma}\delta\gamma=0.\label{eq:QSworklqd}\end{equation}
where $\eta$ is the shear viscosity. This is relevant because the
change in free energy due to the strain is exactly given by the quasistatic
work done and thus the underlying equilibrium free energy of a fluid
does not change with a strain. 

Let us now consider what happens if we subject an initially unstressed
solid, $\gamma=0$, to an infinitesimal change in strain. We assume
that the final strain $\delta\gamma$ is sufficiently small that the
solid responds according to linear elasticity theory. Thus the average
shear stress is related to the zero frequency shear modulus, $G_{0}$,
and the strain by the equation $\left\langle P_{xy}(t)\right\rangle =-G_{0}\gamma(t)$.
The change $\delta\gamma$ may be effected by perturbing the boundary
conditions, 

\begin{eqnarray}
\left\langle \Delta W\right\rangle _{QS} & = & -\lim_{\dot{\gamma}\rightarrow0}\,\dot{\gamma}V\int_{0}^{\delta\gamma/\dot{\gamma}}dt\,\left\langle P_{xy}(t)\right\rangle \nonumber \\
 & = & \lim_{\dot{\gamma}\rightarrow0}VG_{0}\dot{\gamma}^{2}\int_{0}^{\delta\gamma/\dot{\gamma}}tdt=\frac{1}{2}VG_{0}\delta\gamma^{2}.\label{eq:QSworksolid}\end{eqnarray}
Because a solid can support a stress for an indefinite time the underlying
equilibrium free energy will depend upon the change in strain, as
in turn will the partition function, Eq. (\ref{Helmholtz}). Thus
the expression for the equilibrium distribution function will now
explicitly depend upon the strain through the partition function,
$Z(\delta\gamma)$, and so will the equilibrium average of a phase
variable, $\left\langle B(\mathbf{\Gamma})\right\rangle _{\delta\gamma,eq}$.

We wish to calculate (to leading order) the change in the xy-element
of the pressure tensor for a solid subject to a shearing deformation
with a strain $\delta\gamma$. The equilibrium average can be calculated
from the expression, \begin{equation}
\left\langle P_{xy}\right\rangle _{\delta\gamma,eq}=\frac{\int_{D(\delta\gamma)}d\mathbf{\Gamma}P_{xy}(\mathbf{\Gamma})\exp[-\beta(H_{0}(\mathbf{\Gamma})]}{\int_{D(\delta\gamma)}d\mathbf{\Gamma}\exp[-\beta(H_{0}(\mathbf{\Gamma})]}.\label{eq:avPxystraindomain}\end{equation}
In this equation $D(\delta\gamma)$ defines a phase space domain which
is strained an amount $\delta\gamma$, from a reference domain $D(0)$.
This may represent changes in the boundary conditions. The average
$\left\langle \ldots\right\rangle _{0,eq}$ is an equilibrium average
taken over the domain $D(0)$ and the average $\left\langle \ldots\right\rangle _{\delta\gamma,eq}$
is similar but taken over the domain $D(\delta\gamma)$. Because of
the anisotropy of crystals the observed stresses will be strong functions
of the alignment of the crystal relative to the strain or strain rate,
tensor. For simplicity we do not use notation that makes this alignment
explicit.

The transformation between the two domains is given by the equation,\begin{equation}
\mathbf{\Gamma}^{\prime}\equiv\mathbf{\Gamma}-\delta\mathbf{\Gamma}\label{eq:delphasevect}\end{equation}
where $\delta\mathbf{\Gamma}$ is

\begin{equation}
\mathbf{\delta}\mathbf{\Gamma}=\delta\gamma(q_{y1},0,0,q_{y2},0,0,\ldots,q_{yN},0,0,0,0,0,0\ldots,0,0,0).\label{eq:straindiffphase}\end{equation}
We can transform the average Eq. (\ref{eq:avPxystraindomain}) using
the coordinate transformation as\begin{equation}
\left\langle P_{xy}\right\rangle _{\delta\gamma,eq}=\frac{\int_{D(0)}d\mathbf{\Gamma}^{\prime}\left|\frac{\partial\mathbf{\Gamma}}{\partial\mathbf{\Gamma}^{\prime}}\right|P_{xy}(\mathbf{\Gamma}^{\prime}+\delta\mathbf{\Gamma})\exp[-\beta(H_{0}(\mathbf{\Gamma}^{\prime}+\delta\mathbf{\Gamma})]}{\int_{D(0)}d\mathbf{\Gamma}^{\prime}\left|\frac{\partial\mathbf{\Gamma}}{\partial\mathbf{\Gamma}^{\prime}}\right|\exp[-\beta(H_{0}(\mathbf{\Gamma}^{\prime}+\delta\mathbf{\Gamma})]}.\label{eq:Pxy1}\end{equation}
Noting that the Jacobian is unity, that $d\mathbf{\Gamma}^{\prime}$
is a dummy integration variable and expanding $P_{xy}$ and $H_{0}$
to leading orders in $\delta\gamma$ gives\begin{equation}
\left\langle P_{xy}\right\rangle _{\delta\gamma,eq}=\frac{\int_{D(0)}d\mathbf{\Gamma}[P_{xy}(\mathbf{\Gamma})+\delta\mathbf{\Gamma}\cdot\mathbf{\nabla}P_{xy}(\mathbf{\Gamma})]\exp[-\beta([H_{0}(\mathbf{\Gamma})+\delta\mathbf{\Gamma}\cdot\mathbf{\nabla}H_{0}(\mathbf{\Gamma})]]}{\int_{D(0)}d\mathbf{\Gamma}\exp[-\beta([H_{0}(\mathbf{\Gamma})+\delta\mathbf{\Gamma}\cdot\mathbf{\nabla}H_{0}(\mathbf{\Gamma})]]}.\end{equation}
Approximating the exponentials to leading order in $\delta\mathbf{\Gamma}$
gives,\begin{equation}
\left\langle P_{xy}\right\rangle _{\delta\gamma,eq}=\frac{\int_{D(0)}d\mathbf{\Gamma}\left[P_{xy}(\mathbf{\Gamma})+\delta\mathbf{\Gamma}\cdot\mathbf{\nabla}P_{xy}(\mathbf{\Gamma})\right]\left[1-\beta\delta\mathbf{\Gamma}\cdot\mathbf{\nabla}H_{0}(\mathbf{\Gamma})\right]\exp[-\beta(H_{0}(\mathbf{\Gamma})]}{\int_{D(0)}d\mathbf{\Gamma}\exp[-\beta(H_{0}(\mathbf{\Gamma})]-\beta\int_{D(0)}d\mathbf{\Gamma}\delta\mathbf{\Gamma}\cdot\mathbf{\nabla}H_{0}(\mathbf{\Gamma})\exp[-\beta(H_{0}(\mathbf{\Gamma})]]},\label{eq:Pxy3}\end{equation}
and expansion of the denominator, to leading order in $\delta\mathbf{\Gamma}$,
gives,\[
\left\langle P_{xy}\right\rangle _{\delta\gamma,eq}=\frac{\int_{D(0)}d\mathbf{\Gamma}\left[P_{xy}(\mathbf{\Gamma})+\delta\mathbf{\Gamma}\cdot\mathbf{\nabla}P_{xy}(\mathbf{\Gamma})\right]\left[1-\beta\delta\mathbf{\Gamma}\cdot\mathbf{\nabla}H_{0}(\mathbf{\Gamma})\right]\exp[-\beta(H_{0}(\mathbf{\Gamma})]}{\int_{D(0)}d\mathbf{\Gamma}exp[-\beta(H_{0}(\mathbf{\Gamma})]}\]
\begin{equation}
\times\left(1+\frac{\beta\int_{D(0)}d\mathbf{\Gamma}\delta\mathbf{\Gamma}\cdot\mathbf{\nabla}H_{0}(\mathbf{\Gamma})exp[-\beta(H_{0}(\mathbf{\Gamma})]}{\int_{D(0)}d\mathbf{\Gamma}exp[-\beta(H_{0}(\mathbf{\Gamma})]}\right).\label{eq:Pxy4}\end{equation}
Applying the coordinate transformation, Eqs. (\ref{eq:delphasevect})
\& (\ref{eq:straindiffphase}), we see that

\begin{equation}
\delta\mathbf{\Gamma}\cdot\nabla H_{0}(\mathbf{\Gamma})=-\delta\gamma P_{xy}^{\Phi}(\mathbf{\Gamma})V,\label{eq:strainH0}\end{equation}
where $P_{xy}^{\Phi}$ is the configurational component of the xy-element
of the pressure tensor and\begin{equation}
\delta\mathbf{\Gamma}\cdot\mathbf{\nabla}P_{xy}(\mathbf{\Gamma})V=\delta\gamma\sum_{i=1}^{N}\sum_{j=1}^{N}\frac{\partial F_{xi}}{\partial q_{xj}}q_{yi}q_{yj}.\label{eq:strainPxyV}\end{equation}
Substitution into Eq. (\ref{eq:Pxy4}) gives

\begin{eqnarray}
\left\langle P_{xy}\right\rangle _{\delta\gamma,eq}V & = & \left\langle P_{xy}\right\rangle _{0,eq}V\nonumber \\
 &  & +\beta\delta\gamma V^{2}\left[\left\langle P_{xy}P_{xy}^{\Phi}\right\rangle _{0,eq}-\left\langle P_{xy}\right\rangle _{0,eq}\left\langle P_{xy}^{\Phi}\right\rangle _{0,eq}\right]\nonumber \\
 &  & +\delta\gamma\left\langle \sum_{i=1}^{N}\sum_{j=1}^{N}\frac{\partial F_{xi}}{\partial q_{xj}}q_{yi}q_{yj}\right\rangle _{0,eq}+{\cal O}\left(\delta\gamma^{2}\right)\nonumber \\
 & = & \left\langle P_{xy}\right\rangle _{0,eq}V+\beta\delta\gamma V^{2}\left[\left\langle P_{xy}^{2}\right\rangle _{0,eq}-\left\langle P_{xy}\right\rangle _{0,eq}^{2}\right]-\delta\gamma V\left\langle g_{\infty}\right\rangle _{0,eq}+{\cal O}\left(\delta\gamma^{2}\right).\label{eq:Pxyfinalstrain}\end{eqnarray}
Where for simplicity we define the phase function $g_{\infty}$ by
the equation\begin{equation}
g_{\infty}V\equiv\sum_{i=1}^{N}\frac{p_{yi}^{2}}{m_{i}}-\sum_{i=1}^{N}\sum_{j=1}^{N}\frac{\partial F_{xi}}{\partial q_{xj}}q_{yi}q_{yj}\label{eq:def_ginfV}\end{equation}
which implies the zero frequency shear modulus is \begin{equation}
G_{0}=\left\langle g_{\infty}\right\rangle _{0,eq}-\beta V\left\langle \left[P_{xy}-\left\langle P_{xy}\right\rangle _{0,eq}\right]^{2}\right\rangle _{0,eq}.\label{eq:defG0}\end{equation}
The zero frequency shear modulus is thus the sum of a fluctuation
and a nonfluctuating term. The nonfluctuating term (valid at zero
temperature) was given by Born \citep{Born-JCP-1939,Born-1954} in
1939. The derivation of the correct finite temperature result was
first given by Squire et. al. \citep{Squire-P-1969,Hoover-P-1969}
in 1969 and rediscovered in 1986, see ref. \onlinecite{Bavaud-JOSP-1986}.
For a fluid, the sum of these two terms is \emph{exactly} zero since
the shear modulus is zero \citep{Hess-PA-1997}. For a solid these
two terms do not cancel and there is a non-zero shear modulus.

To gain a better understanding of these two terms consider the response
of a system to an impulsive strain rate: $\dot{\gamma}(t)=\delta\gamma\delta(t)$.
From the SLLOD equations of motion Eq. (\ref{sllod}) we see that
for impulsive shear the change in the phase space vector is

\begin{equation}
\mathbf{\delta\Gamma}=\delta\gamma(q_{y1},0,0,q_{y2},0,0,\ldots,q_{yN,0,0},-p_{y1},0,0,-p_{y2},0,0,\ldots,-p_{yN},0,0).\label{eq:slloddifphase}\end{equation}
If we now consider some phase variable $B(\mathbf{\Gamma})$ whose
functional form is not explicitly dependent on the strain, \begin{equation}
B(\mathbf{\Gamma}(0^{+}))=B(\mathbf{\Gamma}(0^{-}))+\nabla B(\mathbf{\Gamma}(0^{-}))\cdot\delta\mathbf{\Gamma}(\delta\gamma)+{\cal O}(\delta\gamma^{2}),\label{eq:sSLLODB}\end{equation}
and substitute $P_{xy}V$ for $B$ we see that \begin{equation}
\left\langle P_{xy}(0^{+})\right\rangle V=\left\langle P_{xy}\right\rangle _{0,eq}V-\delta\gamma V\left\langle g_{\infty}\right\rangle _{0,eq}\equiv\left\langle P_{xy}\right\rangle _{0,eq}V-\delta\gamma G_{\infty}V,\label{eq:Pxyimpulsestrain}\end{equation}
where $\left\langle P_{xy}(t)\right\rangle $ is the nonequilibrium
average taken, at time $t$, in this case at time $t=0^{+}$ which
is directly after the system has been subjected to the impulse. The
nonfluctuating component of the \emph{zero} frequency shear modulus
is in fact the \emph{infinite} frequency shear modulus, $G_{\infty}$,
so that, $G_{0}=G_{\infty}-\beta V\left\langle \left[P_{xy}-\left\langle P_{xy}\right\rangle _{0,eq}\right]^{2}\right\rangle _{0,eq}$.
For all systems (solids or fluids) the infinite frequency shear modulus
is given by Eq. (\ref{eq:def_ginfV}). At infinite frequency there
is no time for the system to recognize whether it is a fluid or a
solid. In a fluid, \emph{and only in a fluid}, the zero frequency
shear modulus is zero and hence we get a second exact expression for
the infinite frequency shear modulus that is only valid for fluids:
$G_{\infty}^{F}=\beta V\left\langle \left[P_{xy}-\left\langle P_{xy}\right\rangle _{0,eq}\right]^{2}\right\rangle _{0,eq}$.
Here the superscript \emph{$F$} indicates that this expression is
only valid for fluids. This latter expression is familiar to those
acquainted with the Green-Kubo expressions for the frequency dependent
shear viscosity of fluids. 

Thus regardless of what state of matter we are dealing with

\begin{equation}
\beta V\left\langle \left[P_{xy}-\left\langle P_{xy}\right\rangle _{0,eq}\right]^{2}\right\rangle _{0,eq}=G_{\infty}-G_{0}.\label{eq:Gdifference}\end{equation}
However the difference between a solid and a fluid is whether the
zero frequency modulus is zero as in a fluid or a positive number
as in a solid.

It is important to remember that this perfect mathematical cancellation
between the fluctuation term and the nonfluctuating term in fluids
is no simple mathematical identity. Consider two systems with identical
Hamiltonians, densities and equations of motion. The only difference
between the two systems is their temperature. One is in the solid
state phase and the other is in the liquid phase. You can transform
between the two states by simply changing the temperature. Yet in
the liquid the cancellation is perfect whereas in the solid it is
not. This cancellation is a symmetry that is particular to the fluid
state. 

The standard derivations of linear response theory assume that the
underlying equilibrium distribution function does not change with
strain. Clearly if we wish to treat a solid phase we must account
for the effect of the underlying equilibrium distribution function
depending on the strain.

\subsection{Linear response to shear for systems that are initially at equilibrium}

In this section we will consider the linear change in the stress in
response to a small applied strain, firstly for an ordinary fluid,
then a crystal which is initially unstrained and finally for a crystal
with an initial strain that is not zero.

Linear response theory is described in terms of a field and a conjugate
flux. For the special case where both the field and the flux, which
are vectors, have a common direction, only their magnitudes are relevant
and we may define the flux, $J$, as,\begin{equation}
J(\mathbf{\Gamma})\equiv\frac{\dot{Q}(\mathbf{\Gamma})-\dot{H}_{0}(\mathbf{\Gamma})}{VF_{e}(t)},\label{flux-def}\end{equation}
where $V$ is the system volume, $F_{e}$ is the magnitude of the
field which appears in Eqs. (\ref{EOM}) and $\dot{Q}$ is the rate
at which heat is exchanged with the synthetic thermostat. To leading
order in the field $\left\langle \dot{Q}(\mathbf{\Gamma})-\dot{H}_{0}(\mathbf{\Gamma})\right\rangle ={\cal O}(F_{e}^{2})$
and thus as the field approaches zero so does the flux, $J(\mathbf{\Gamma})={\cal O}(F_{e})$.

\subsubsection{Equilibrium fluid}

Consider a fluid in the isokinetic ensemble, Eq. (\ref{Eq-dist}),
which is initially in equilibrium and then perturbed by an external
field at time $t=0$. Linear response theory gives, \citep{evans-and-morriss-book},\begin{equation}
\left\langle B(t)\right\rangle =\left\langle B\right\rangle _{eq}-\beta V\int_{0}^{t}ds\,\left\langle J(-s)B(0)\right\rangle _{eq}F_{e}(t-s),\label{LRT-fluid}\end{equation}
where $B$ is some arbitrary phase variable, $\left\langle J(-s)B(0)\right\rangle _{eq}=\left\langle J\boldsymbol{(}\boldsymbol{\Gamma}(-s)\boldsymbol{)}B(\boldsymbol{\Gamma})\right\rangle _{eq}$
and $\boldsymbol{\Gamma}(-s)$ is the point in phase space, such that
if we start at it and run the equations of motion forward in time
(with $F_{e}=0$ because the average is an equilibrium one) we arrive
at the point $\boldsymbol{\Gamma}$ at time $0$. Here we are interested
in planar shear with $B(\boldsymbol{\Gamma})=J(\boldsymbol{\Gamma})=P_{xy}(\boldsymbol{\Gamma})$,
$\left\langle P_{xy}\right\rangle _{eq}=0$ and $F_{e}(t)=\dot{\gamma}(t)$.
So we obtain\begin{equation}
\left\langle P_{xy}(t)\right\rangle =-\beta V\int_{0}^{t}ds\,\left\langle P_{xy}(-s)P_{xy}(0)\right\rangle _{eq}\dot{\gamma}(t-s).\label{eq:GKPxy}\end{equation}

\subsubsection{Initially Unstrained crystal}

For the case of a crystal, the underlying free energy changes with
the strain due to the change in the boundary conditions and a stress
can be supported indefinitely. Of course there are other processes
than just planar shear which could result in this behaviour and so
we will represent the change in the free energy using the arbitrary
parameter $\lambda$. In the case of planar shear we will have $\lambda=\gamma$.
We have recently given a generalisation of linear response theory
for such a case where a system may be simultaneously subject to a
dissipative field, $F_{e}$, and a parametric change, $\lambda(t)$,
to its equilibrium state \citep{Williams-PRE-2008}. In this paper
we proved that to linear order in an arbitrary dissipative field and
parameter the average linear response of a phase variable $B(\mathbf{\Gamma},\lambda)$
that may depend on the parameter is,\begin{eqnarray}
\left\langle B(t)\right\rangle _{\lambda(t)} & = & \left\langle B\right\rangle _{\lambda(t),eq}-\beta V\int_{0}^{t}ds\,\left\langle J(-s)B(0)\right\rangle _{\lambda(0),eq}F_{e}(t-s)\nonumber \\
 &  & -\beta\int_{0}^{t}ds\,\left[\frac{\partial A(\lambda(0))}{\partial\lambda}\left\langle B(\mathbf{\Gamma},\lambda(s))\right\rangle _{\lambda(0),eq}-\left\langle \frac{\partial H(\mathbf{\Gamma}(-s),\lambda(0)}{\partial\lambda}B(\mathbf{\Gamma},\lambda(s))\right\rangle _{\lambda(0),eq}\right]\dot{\lambda}(t-s).\label{eq:LRTDissipation&parametric}\end{eqnarray}

We choose to set the parameter as the strain and the dissipative field
as the strain rate. In this case the Hamiltonian, $H_{0}$, and the
phase function $B(\mathbf{\Gamma})$, have no explicit dependence
on the strain or strain rate, but averages will still be dependent
on these parameters via the boundary conditions. Applying Eq(\ref{eq:LRTDissipation&parametric})
to this situation gives\begin{eqnarray}
\left\langle B(t)\right\rangle _{\gamma(t)} & = & \left\langle B\right\rangle _{\gamma(t),eq}-\beta V\int_{0}^{t}ds\,\left\langle P_{xy}(-s)B(0)\right\rangle _{\gamma(0),eq}\dot{\gamma}(t-s)\nonumber \\
 &  & -\beta\frac{\partial A(\gamma(0))}{\partial\gamma}\left\langle B\right\rangle _{\gamma(0),eq}\int_{0}^{t}ds\,\dot{\gamma}(t-s),\label{Eq-gen-LRT}\end{eqnarray}
where $\left\langle B(t)\right\rangle _{\gamma(t)}$ is the value
of the nonequilibrium average at time $t$. 

Let us now consider the example of a planar shear impulse again, and
choose $B(\boldsymbol{\Gamma})=J(\boldsymbol{\Gamma})=P_{xy}(\boldsymbol{\Gamma})$,
$\gamma(t)=0\;\forall\; t<0$ and $\dot{\gamma}(t)=\gamma_{1}\delta(t)$.
Because the initial stress is zero we will have $\left\langle P_{xy}\right\rangle _{\gamma(t),eq}=0\;\forall\; t<0$.
We consider the terms on the right hand side of Eq. (\ref{Eq-gen-LRT})
in turn. The first term is easily seen to be\begin{equation}
\left\langle P_{xy}\right\rangle _{\gamma_{1},eq}=-G_{0}\gamma_{1}+{\cal O}(\gamma_{1}^{3}).\label{eq-solid-strain0}\end{equation}
The second term is easily evaluated as $-\beta V\gamma_{1}\left\langle P_{xy}\boldsymbol{(}\boldsymbol{\Gamma}(-t)\boldsymbol{)}P_{xy}(\boldsymbol{\Gamma})\right\rangle _{0,eq}$.
Lastly we need to calculate the change in the equilibrium free energy
caused by the strain. We note that Eq. (\ref{Helmholtz}) relates
the free energy to the partition function. The change in the partition
function is easily computed as

\[
Z(N,V,T,\gamma_{1})=\int_{D(0)}d\mathbf{\Gamma}\exp[-\beta(H_{0}(\mathbf{\Gamma})]-\beta\int_{D(0)}d\mathbf{\Gamma}\delta\mathbf{\Gamma}\cdot\mathbf{\nabla}H_{0}(\mathbf{\Gamma})\exp[-\beta(H_{0}(\mathbf{\Gamma})]]\]
\begin{equation}
=Z(N,V,T,\gamma(0^{-}))\left[1+\beta V\gamma_{1}\left\langle P_{xy}^{\Phi}\right\rangle _{\gamma(0^{-}),eq}\right],\label{eq:partitionstrain1}\end{equation}
and the free energy is thus

\begin{equation}
A(N,V,T,\gamma_{1})=A(N,V,T,\gamma(0^{-}))-\gamma_{1}V\left\langle P_{xy}^{\Phi}\right\rangle _{0}.\label{eq:Helmholtzstrain1}\end{equation}
Given the reference domain $D(0)$ has zero strain and zero stress,
$\left\langle P_{xy}\right\rangle _{0,eq}=0$, the change in the free
energy will be $\mathcal{O}(\gamma^{2})$ and may be ignored in Eq.
(\ref{Eq-gen-LRT}). We now use Eqs. (\ref{eq:defG0}), (\ref{eq-solid-strain0})
\& (\ref{Eq-gen-LRT}) to obtain the response to the impulse, $\dot{\gamma}(t)=\gamma_{1}\delta(t)$
\begin{eqnarray}
\left\langle P_{xy}(t)\right\rangle  & = & -\gamma_{1}\left\langle g_{\infty}\right\rangle _{0,eq}+\beta V\gamma_{1}\left\langle P_{xy}^{2}\right\rangle _{0,eq}\nonumber \\
 &  & -\beta V\gamma_{1}\left\langle P_{xy}(-t)P_{xy}(0)\right\rangle _{0,eq}.\label{LRT-crystal-gam0}\end{eqnarray}
At time $t=0^{+}$, Eq. (\ref{LRT-crystal-gam0}) coincides with the
stress that one would calculate for a sudden impulse (ie the response
is given by the infinite frequency shear modulus alone). At very long
times, $t\rightarrow\infty$ (where the autocorrelation function fully
decays), Eq. (\ref{LRT-crystal-gam0}) reduces to that given by the
zero frequency shear modulus Eq. (\ref{eq:defG0}). At first sight,
Eq.(\ref{Eq-gen-LRT}) looks paradoxical. The first term on the right
hand side is the correct long time answer for any changes in the shear
rate that are completed before time \emph{t}. Since that first term
has no memory it looks as though it gives the infinite frequency response.
However, as this example has just proved, this is not the case. 

If we apply Eq. (\ref{LRT-crystal-gam0}) to a fluid, which must have
$\left\langle g_{\infty}\right\rangle _{eq}=\beta V\left\langle P_{xy}^{2}\right\rangle _{eq}$,
then Eq. (\ref{LRT-crystal-gam0}) will be compatible with Eq. (\ref{eq:GKPxy}).

\subsubsection{Initially Strained crystal}

For a crystal which is initially strained by an amount $\gamma_{0}=\gamma(0^{-})$,
but that is none the less in equilibrium, Eq. (\ref{eq-solid-strain0})
becomes\begin{equation}
\left\langle P_{xy}\right\rangle _{\gamma_{0}+\gamma_{1},eq}=\left\langle P_{xy}\right\rangle _{\gamma_{0},eq}-G_{0}(\gamma_{0})\gamma_{1}+{\cal O}(\gamma_{1}^{2}),\label{eq:solidstrainstrain}\end{equation}
and for the change in free energy we have \begin{equation}
\frac{\partial A(\gamma_{0})}{\partial\gamma}=-V\left\langle P_{xy}\right\rangle _{\gamma_{0},eq}.\label{eq:DAstrainstrain}\end{equation}
We now use Eq. (\ref{Eq-gen-LRT}) and consider the response to an
impulsive change in the strain rate, $\dot{\gamma}(t)=\delta(t)\gamma_{1}$.
We see that\begin{eqnarray}
\left\langle P_{xy}(t)\right\rangle  & = & \left\langle P_{xy}\right\rangle _{\gamma_{0},eq}-G_{0}(\gamma_{0})\gamma_{1}-\beta V\gamma_{1}\left\langle \Delta P_{xy}\boldsymbol{(}\boldsymbol{\Gamma}(-t)\boldsymbol{)}\Delta P_{xy}(\boldsymbol{\Gamma})\right\rangle _{\gamma_{0},eq}\nonumber \\
 & = & \left\langle P_{xy}\right\rangle _{\gamma_{0},eq}-\gamma_{1}\left\langle g_{\infty}\right\rangle _{\gamma_{0},eq}+\beta V\gamma_{1}\left\langle \Delta P_{xy}^{2}\right\rangle _{\gamma_{0},eq}-\beta V\gamma_{1}\left\langle \Delta P_{xy}\boldsymbol{(}\boldsymbol{\Gamma}(-t)\boldsymbol{)}\Delta P_{xy}(\boldsymbol{\Gamma})\right\rangle _{\gamma_{0},eq},\label{LRT-crystal-gam1}\end{eqnarray}
where $\Delta P_{xy}(\boldsymbol{\Gamma})=P_{xy}(\boldsymbol{\Gamma})-\left\langle P_{xy}\right\rangle _{\gamma_{0},eq}$.
Again we see that this equation gives the initial response to the
impulse,\begin{equation}
\left\langle P_{xy}(0^{+})\right\rangle =\left\langle P_{xy}\right\rangle _{\gamma_{0},eq}-\gamma_{1}\left\langle g_{\infty}\right\rangle _{\gamma_{0},eq},\label{eq:Pxy-crystal-gam1}\end{equation}
at time $0^{+}$ and as $t\rightarrow\infty$ it decays to the equilibrium
value,\begin{equation}
\left\langle P_{xy}\right\rangle _{\gamma_{0}+\gamma_{1},eq}=\left\langle P_{xy}\right\rangle _{\gamma_{0},eq}-\gamma_{1}\left\langle g_{\infty}\right\rangle _{\gamma_{0},eq}+\beta V\gamma_{1}\left\langle \Delta P_{xy}^{2}\right\rangle _{\gamma_{0},eq}.\end{equation}

\begin{equation}
=\left\langle P_{xy}\right\rangle _{\gamma_{0},eq}-G_{0}(\gamma_{0})\gamma_{1}.\label{eq:Pxy-crystal-eqm}\end{equation}
For the case where $\gamma_{0}=0$ we have $\left\langle P_{xy}\right\rangle _{\gamma_{0},eq}=0$
and $\Delta P_{xy}(\boldsymbol{\Gamma})=P_{xy}(\boldsymbol{\Gamma})$
and Eq. (\ref{LRT-crystal-gam1}) reduces to Eq. (\ref{LRT-crystal-gam0}).
We reiterate that the values for the various elastic moduli will of
course depend on the alignment of the crystal relative to the strain
rate tensor. For a crystal the elastic modulus is in fact a $4^{th}$
rank polar tensor.

\subsection{Oscillatory Planar Shear}

We now consider the case of oscillatory planar shear applied to a
crystal, by using Eq. (\ref{Eq-gen-LRT}) to calculate the response
in the stress $B(\mathbf{\Gamma})=P_{xy}$ to an applied strain of
the form,\begin{equation}
\gamma(t)=\gamma_{0}\sin(\omega t)=-\Re\left\{ i\gamma_{0}e^{i\omega t}\right\} .\label{eq:defstrainosc}\end{equation}
The response to the oscillatory strain will become sinusoidal after
the decay of initial transients and is often expressed in terms of
the storage (or real) $\tilde{G}_{R}$ and loss (or imaginary) $\tilde{G}_{I}$
shear moduli\begin{equation}
\lim_{t\rightarrow\infty}\left\langle P_{xy}(t)\right\rangle =\Re\left\{ i\gamma_{0}\tilde{G}(\omega)e^{i\omega t}\right\} \label{eq:PxyXtalConstit}\end{equation}
where $\tilde{G}(\omega)=\widetilde{G}_{R}(\omega)+i\widetilde{G}_{I}(\omega)$.
This quantity is related to the complex frequency dependent shear
viscosity by the equation, \begin{equation}
\tilde{G}(\omega)=i\omega\tilde{\eta}(\omega),\label{viscosity-def}\end{equation}
where $\tilde{\eta}(\omega)=\widetilde{\eta}_{R}(\omega)-i\widetilde{\eta}_{I}(\omega)$
and thus $G_{0}=\tilde{G}(\omega=0)$ and $G_{\infty}=\underset{\omega\rightarrow\infty}{lim}\tilde{G}(\omega)$.

The applied field is $F_{e}=\dot{\gamma}(t)=\omega\gamma_{0}\cos(\omega t)$,
the change in free energy for a crystal is given by $dA/d\gamma=-\left\langle P_{xy}\right\rangle _{0,eq}=0$
and the flux is $J(\mathbf{\Gamma})=P_{xy}(\mathbf{\Gamma})$. Using
Eq. (\ref{Eq-gen-LRT}) we have\begin{equation}
\left\langle P_{xy}(t)\right\rangle =\left\langle P_{xy}\right\rangle _{\gamma(t),eq}-\beta V\omega\gamma_{0}\int_{0}^{t}ds\,\left\langle P_{xy}(-s)P_{xy}(0)\right\rangle _{\gamma(0),eq}\cos(\omega(t-s))\label{eq:memoryfn 1}\end{equation}
which, using a trigonometric identity and Eq. (\ref{eq-solid-strain0}),
gives\begin{eqnarray}
\left\langle P_{xy}(t)\right\rangle  & = & -G_{0}\gamma_{0}\sin(\omega t)\label{eq:memoryfn 2}\\
 &  & -\beta V\omega\gamma_{0}\sin(\omega t)\int_{0}^{t}ds\,\left\langle P_{xy}(-s)P_{xy}(0)\right\rangle _{\gamma(0),eq}\sin(\omega s)\nonumber \\
 &  & -\beta V\omega\gamma_{0}\cos(\omega t)\int_{0}^{t}ds\,\left\langle P_{xy}(-s)P_{xy}(0)\right\rangle _{\gamma(0),eq}\cos(\omega s).\nonumber \end{eqnarray}
Combining this with Eq. (\ref{eq:PxyXtalConstit}) we obtain, \begin{eqnarray}
\widetilde{G}_{R}(\omega) & = & G_{0}+\beta V\omega\int_{0}^{\infty}ds\,\sin(\omega s)C(s)\nonumber \\
\widetilde{G}_{I}(\omega) & = & \beta V\omega\int_{0}^{\infty}ds\,\cos(\omega s)C(s),\label{G-LRT-xtal}\end{eqnarray}
where the correlation function is given by \begin{equation}
C(s)=\left\langle P_{xy}(-s)P_{xy}(0)\right\rangle _{eq,0}.\label{eq:def-Calpha-xtal}\end{equation}
Using Eqs.(\ref{viscosity-def}) and (\ref{G-LRT-xtal}) we see that,
$\widetilde{\eta}_{R}(\omega)=\widetilde{G}_{I}(\omega)/\omega$ and
$\widetilde{G}_{R}(\omega)=G_{0}+\omega\widetilde{\eta}_{I}(\omega)$.\begin{eqnarray}
\widetilde{\eta}_{R}(\omega) & = & \beta V\int_{0}^{\infty}ds\,\cos(\omega s)C(s)\nonumber \\
\widetilde{\eta}_{I}(\omega) & = & -G_{0}/\omega+\beta V\int_{0}^{\infty}ds\,\sin(\omega s)C(s).\label{eta-LRT-xtal}\end{eqnarray}

If we now ask what memory function $\eta(t)$, \begin{equation}
\left\langle P_{xy}(t)\right\rangle \equiv-\int_{0}^{t}ds\,\eta(t-s)\dot{\gamma}(s),\label{eq:memory function def}\end{equation}
generates this spectrum for the frequency dependent shear viscosity,
we see that it is,\begin{equation}
\eta(t)=G_{0}+\beta VC(t),\; t>0;\label{eq:memory function xtal}\end{equation}
\[
\eta(t)=-G_{0},\; t<0.\]

A double sided Fourier transform of this function gives Eq.(\ref{eta-LRT-xtal}).
The memory function for the zero frequency elastic response must be
odd in time because a constant strain is an even function of time
while a constant strain \emph{rate} is of course odd. For linear elasticity
and linear viscosity to both apply to the system, both the strain
and the strain rate must be small at all frequencies including near
zero.

\section{Simulation, Results and Discussion}

\subsection{Simulation Details}

To test the theory we used both equilibrium time correlation data,
obtained using Eqs. (\ref{EOM}) with $\mathbf{F}_{e}=0$ and nonequilibrium
molecular dynamics (NEMD) data, obtained from Eqs. (\ref{sllod})
from oscillatory strain simulations. The simulations used $N=108$
particles with periodic boundary conditions to model a perfect crystal
at a finite temperature. Because we model a single crystal with no
defects there will be no long range stresses and no large system size
effects. We know from the pioneering studies of a realistic potential
for argon by Barker et. al. \citet{Barker-MP-1971} that for perfect
crystals, 108 atoms is sufficient for quantitative agreement with
experiment.

An equilibrium face centre cubic (FCC) crystal was formed as a cubic
array of periodic molecular dynamics unit cells. The FCC crystal is
commensurate with the simulation cell and the edge of the cube is
in the (1,0,0) direction of the crystal. A pairwise additive WCA potential,\begin{eqnarray}
u_{ij}(r_{ij}) & = & 4\epsilon\left[\left(\frac{\sigma}{r_{ij}}\right)^{12}-\left(\frac{\sigma}{r_{ij}}\right)^{6}+\frac{1}{4}\right]\:\forall\: r_{ij}<2^{\nicefrac{1}{6}}\sigma\nonumber \\
u_{ij}(r_{ij}) & = & 0\:\forall\: r_{ij}\geqslant2^{\nicefrac{1}{6}}\sigma,\label{def-WCA}\end{eqnarray}
was used for the interaction between the particles. The energy unit
is $\epsilon$, the length unit is $\sigma$, and the time unit is
$\sqrt{m\sigma^{2}/\epsilon}$ where $m$ is the mass. The system
has a number density of $\rho=N\sigma^{3}/V=1.15$, which results
in an equilibrium crystal at the two temperatures studied, $T=0.5$
and $T=2.5$ $(\epsilon/k_{B})$.

For the nonequilibrium simulations eight different frequencies where
simulated. For those with a frequency of $\omega=2.513$ or higher,
the maximum amplitude of the strain (as defined in Eq. (\ref{eq:defstrainosc}))
was $\gamma_{0}=0.025$, for the lower frequencies $\gamma_{0}=0.08$
was used. As the frequency is lowered, with fixed maximum strain amplitude,
the signal to noise ratio for the loss component of the shear modulus
deteriorates because the strain rate goes towards zero. The larger
amplitude used at low frequencies helped alleviate this problem a
little.

The equations of motion were integrated using a fourth order Runge-Kutta
integrator. A time step of $dt=0.002$ was used for all simulations
except for the NEMD simulations at the highest frequency where a time
step of $dt=0.001$ was used. All the simulations, both NEMD and equilibrium,
were repeated 1000 times to reduce the statistical uncertainties.
At the highest frequency the NEMD simulations were given at least
6.3 time units to relax to the periodic state and for the lowest frequency
this was extended to 2000 time units. For all but the lowest three
frequencies the NEMD simulations were run for 10 periods to obtain
the response. At the lowest frequency the NEMD simulations were run
for only 2 periods. The longest duration used for producing data from
the equilibrium simulations was 800 time units.

\subsection{Simulation Results and Discussion}

\subsubsection{Equilibrium data}

To calculate the response of the system using Eqs. (\ref{G-LRT-xtal})
or (\ref{eta-LRT-xtal}) we need to first determine the equilibrium
correlation function, Eq. (\ref{eq:def-Calpha-xtal}), and the zero
frequency modulus, Eqs. (\ref{eq:def_ginfV}) \& (\ref{eq:defG0}),
using data from equilibrium simulations. %
\begin{figure}
\includegraphics[scale=2]{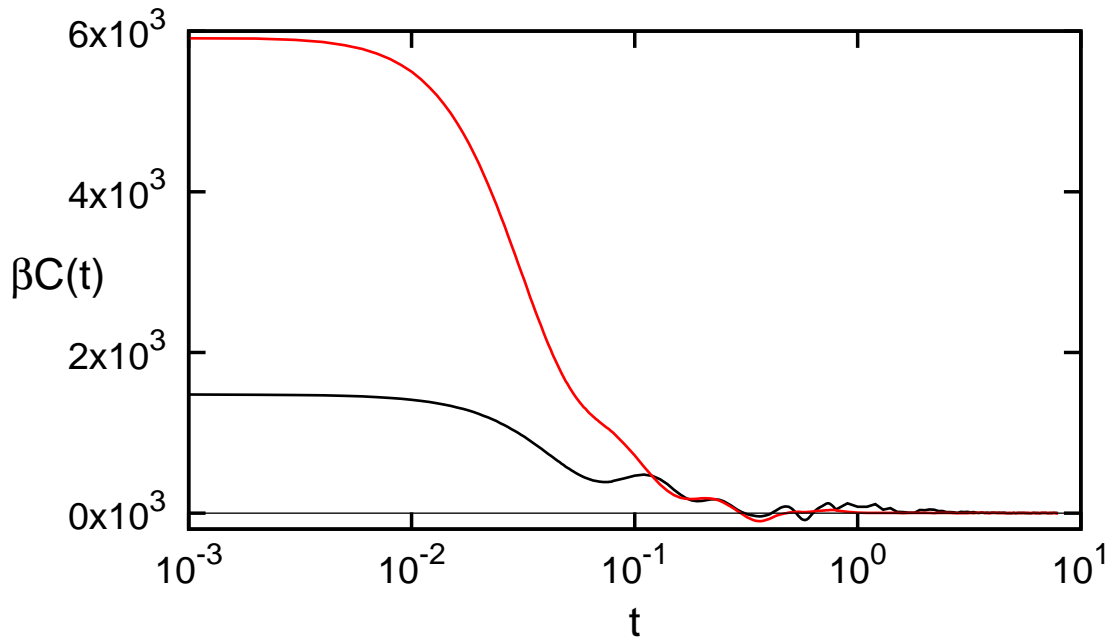}

\caption{The stress correlation function for the FCC crystal at temperatures
of $T=0.5$ and $T=2.5$. Plotted is the function $\beta C(t)=\beta\left\langle P_{xy}(t)P_{xy}(0)\right\rangle $.
Note that the time axis is logarithmic and that the curve with the
higher initial value (close to $6\times10^{3}$) corresponds to the
temperature of $T=2.5$. \label{fig:C(t)}}

\end{figure}
The equilibrium correlation functions, normalised by the temperature,
can be seen in Fig. \ref{fig:C(t)} for the temperatures of $T=0.5$
and $T=2.5$. At zero delay time, $t=0$, the height of the the correlation
function for the higher temperature is approximately a factor of $4$
larger than that for the lower temperature. The viscosity is given
by the area under the curve, $\lim_{\omega\rightarrow0}\tilde{\eta}_{R}(\omega)\equiv\tilde{\eta}_{R}(0^{+})=\beta V\int_{0}^{\infty}ds\, C(s)$,
and as it turns out the values obtained for the two different temperatures
are very similar. For the temperature of $T=2.5$ we have $\tilde{\eta}_{R}(0^{+})=283$
and for the the temperature of $T=0.5$ we have $\tilde{\eta}_{R}(0^{+})=222$.
Although the area under the curve for the higher temperature appears
much larger in Fig. \ref{fig:C(t)}, by noting the logarithmic time
axis it can be seen that the difference decays very rapidly. This
is the reason the values are relatively close to each other. The limiting
viscosity for the crystal increases, rather weakly, with temperature.
This is in contrast to a liquid where the viscosity decreases with
increasing temperature, often strongly, but is similar to a dilute
gas where the viscosity also increases with temperature. 

One cannot measure the viscosity of a crystal at zero frequency by
subjecting it to an unbounded strain. At zero frequency, for large
enough strain, a crystal will exhibit a nonlinear response, undergo
plastic deformation, cleavage or fracture. The viscosity we calculate
is the limiting zero frequency shear viscosity, $\lim(\omega\rightarrow0)$.
If we subject the crystal to an oscillating strain, the viscosity
characterises the dissipation of energy in the low frequency limit. 

An alternative way to measure the limiting zero frequency shear viscosity
is to subject a crystal to a fixed but very small strain rate for
a limited period of time $t_{l}$ which is inversely proportional
to the strain rate, $t_{l}=\gamma_{m}/\dot{\gamma}$. If we set the
maximum strain, $\gamma_{m}$ to be $\gamma_{m}<\sim0.1$ - according
to Lindemann's criterion, then we will not cleave or otherwise damage
the crystal and we will remain in the linear response regime for both
the shear rate which is always very small and for elastic deformation.
As the strain rate decreases towards zero the amount of time we may
strain the crystal diverges to infinity. So even though the maximum
strain the crystal is limited $(\gamma<\gamma_{m})$ as the strain
rate is reduced we have ever more time available for the shearing
system to relax to a nonequilibrium. The shear viscosity of this limiting
steady state is what we call the limiting shear viscosity of a crystal.
Of course as the strain rate is lowered the deteriorating signal to
noise ratio demands that the size of the system, or the number of
times the shearing protocol is repeated must be increased. This may
not be a practical way to measure the limiting viscosity of a solid.

\subsubsection{Frequency Dependent Modulus and Viscosity at $T=0.5$}

\begin{figure}
\includegraphics[scale=2]{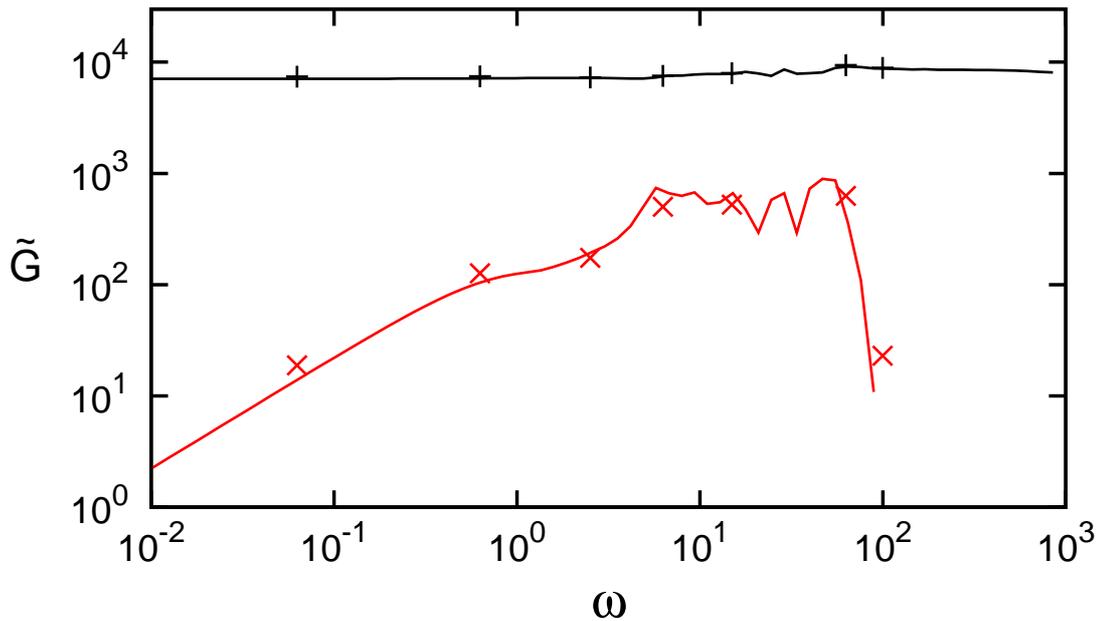}

\caption{The storage $\widetilde{G}_{R}(\omega)$ and loss $\widetilde{G}_{I}(\omega)$
moduli for the temperature of $T=0.5$. The storage modulus $\widetilde{G}_{R}(\omega)$
is the symbols ($+$) while the loss modulus is the symbols ($\times$).
The solid curves are obtained from linear response theory using data
obtained from the equilibrium simulations while the symbols were obtained
directly from the nonequilibrium molecular dynamics simulations.\label{fig:G-star}}

\end{figure}

We examine the case for the FCC crystal at a temperature of $T=0.5$
in more detail. The correlation function $C(t)=\left\langle P_{xy}(t)P_{xy}(0)\right\rangle $
was calculated from equilibrium simulations as discussed above, and
Fourier transformed, according to Eqs. (\ref{G-LRT-xtal}), using
the first order Filon's quadrature given in the Appendix, to obtain
the storage and loss moduli. The value for $G_{0}$ was obtained from
the equilibrium simulations using Eqs. (\ref{eq:def_ginfV}) \& (\ref{eq:defG0}).
We then performed nonequilibrium simulations at various frequencies,
and applied a least squares fit of a sinusoidal function to the response
allowing us to obtain estimates of $\widetilde{G}$ at a small number
of distinct frequencies. The results of this are shown in Fig. \ref{fig:G-star}.
It can be seen that the agreement between the two data sets is very
good. This confirms the correctness of our theoretical expressions
for the frequency dependent elastic moduli. In general the low frequency
$\widetilde{G}_{I}$ data is difficult (or computationally expensive)
to obtain reliably due to the small amplitude of the strain rate.
As mentioned above the strain amplitude is fixed, and therefore the
strain rate becomes very small at low frequencies with $\dot{\gamma}\sim\omega$.
It can be seen that the low frequency data, from the transformed $C(t)$,
decays as $\lim_{\omega\rightarrow0}\widetilde{G}_{I}(\omega)\propto\omega$
(this results in a gradient of unity for $\ln[\widetilde{G}_{I}]$
vs $\ln[\omega]$ at low frequencies, see Fig. \ref{fig:G-star}).
This is obvious from Eq. (\ref{G-LRT-xtal}), upon taking the small
angle approximation $\cos(\omega s)=1+{\cal O}(\omega^{2})$, we see
that $\lim_{\omega\rightarrow0}\widetilde{G}_{I}(\omega)=\omega\widetilde{\eta}_{R}(0^{+})$. 

\begin{figure}
\includegraphics[scale=2]{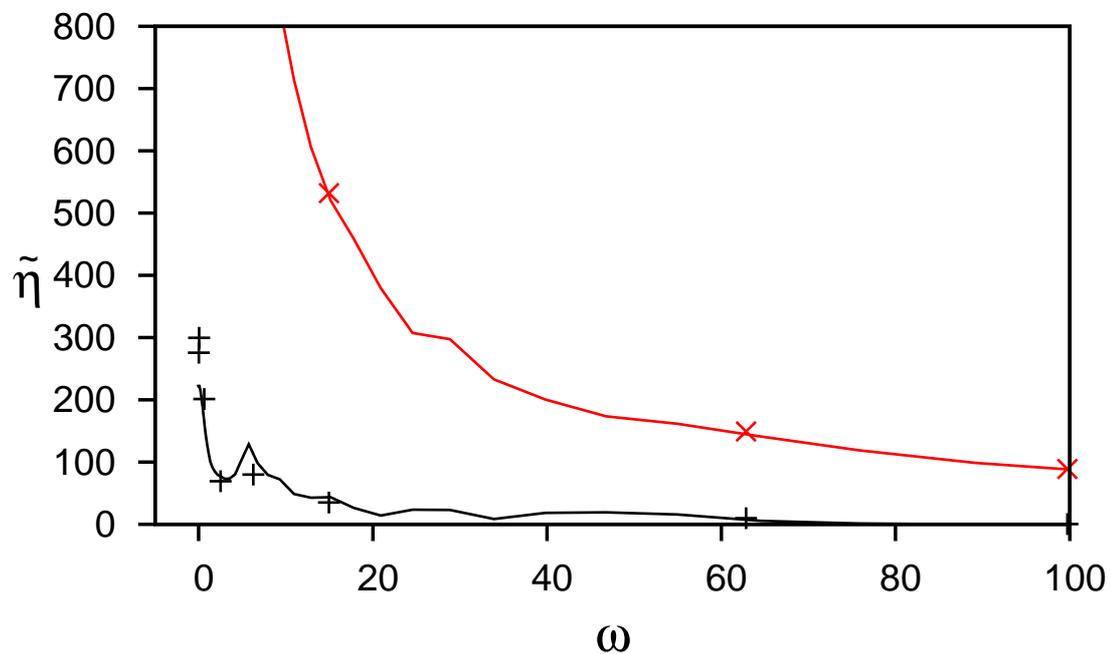}

\caption{Frequency dependent viscosity for the crystal at \emph{T} = 0.5. The
symbols ($+$) represent the real part of the viscosity $\widetilde{\eta}_{R}(\omega)$
and the symbols ($\times$) represent the imaginary part $\widetilde{\eta}_{I}(\omega)$.
The solid lines are from the linear response theory and the symbols
are from NEMD simulations. The symbols for the real part at the lowest
two frequencies are not very accurate. \label{fig:visco}}

\end{figure}
In Fig. \ref{fig:visco} the same data is shown for the complex frequency
dependence of the shear viscosity. The real part of the viscosity,
$\tilde{\eta}_{R}(\omega)$, converges to the finite value at zero
frequency which is given by the area under the correlation function
shown in Fig. \ref{fig:C(t)}. What is distinctly different in this
graph, relative to equivalent data for a typical fluid, is the behaviour
of the imaginary part of the viscosity $\tilde{\eta}_{I}(\omega)$.
We see that this quantity diverges, apparently to infinity, as the
frequency approaches zero. This is due to a solid having a nonzero
value for the zero frequency shear modulus $\widetilde{G}(0)$. In
a fluid the zero frequency value of the imaginary part of the shear
viscosity is zero. The distinctive characteristic of the frequency
dependent viscosity of a solid phase, relative to a fluid phase, is
the contrasting behaviour of the imaginary part of the viscosity as
the frequency approaches zero. For solids the imaginary part diverges
to infinity while in fluids it decays to zero.

\section{Conclusions}

We have derived a set of theoretical expressions for the linear viscoelastic
properties of crystals. Computer simulations have been carried out
which compare the results of direct nonequilibrium molecular dynamics
calculations for these properties, with the linear response theory
expressions calculated using equilibrium simulation data. The agreement
between these two sets of results confirms the correctness of our
expressions for the linear response.

A glass is often defined as a supercooled liquid with a shear viscosity
that is greater than $10^{13}$ poise \citet{Debenedetti-Nat-01}.
In our units, assuming that our potential gives a very approximate
model for argon, this would correspond to a viscosity of approximately
$10^{16}$. However we should point out that when this statement is
made it is also assumed that the shear modulus of the supercooled
liquid/glass is zero. Once we enter the glass phase the shear modulus
is actually nonzero and then according to our theory the Green-Kubo
expression for the shear viscosity changes reducing the magnitude
of the shear viscosity somewhat. 

If the systems we studied here are taken to represent argon we have
shown via equilibrium and nonequilibrium molecular dynamics calculations
that the limiting zero frequency shear viscosity $\widetilde{\eta}_{R}(0^{+})$
of an argon crystal is only two orders of magnitude greater than liquid
argon at its triple point. {[}In the units used in this paper the
shear viscosity of triple point liquid argon is approximately 3.5.{]}
So in contrast to a glass, the crystalline state has no \emph{anomalously}
high shear viscosity. Although we have only calculated this viscosity
for a single relative alignment between the crystal axes and the strain
rate tensor, we do not expect that varying this alignment would lead
to an increase in viscosity of many orders of magnitude. We know of
no other work, experimental or theoretical, that has calculated the
limiting shear viscosity of a crystal. As far as we are aware, all
such work refers to the real and imaginary parts of the shear modulus.

The crystal we have studied is a soft inert gas crystal. It is interesting
to speculate about the comparative values of the limiting shear viscosities
of ionic or covalent crystals. Somewhat counter intuitively these
types of crystals could have limiting viscosity values that at room
temperature, may be lower than that of soft inert gas crystals. What
is important in increasing the viscosity of crystals is anharmonicity
in the nearest neighbour forces. Strong, high Q, harmonic crystals
can be expected to have low shear viscosities.

Our work also points out the distinctiveness of glassy systems. Glassy
systems \emph{by definition} exhibit an anomalously high viscosity.
This is quite different from the behaviour of crystalline solids and
of the liquids from which a glass may be formed.

We also see an interesting set of contrasting qualitative behaviour
for the temperature dependence of the limiting shear viscosity. Gases
exhibit a positive temperature coefficient, liquids have a negative
temperature coefficient while crystals again exhibit a positive temperature
coefficient for the limiting shear viscosity.

The zero frequency shear modulus is profoundly different between a
crystal and a fluid. For fluids the shear modulus is \emph{precisely}
zero whereas in any solid, including crystals, the shear modulus is
nonzero. As Max Born stated in 1939, reference \onlinecite{Born-JCP-1939}, ``..there can be no ambiguity
in the definition of, or the criterion for, melting. The difference
between a solid and a liquid is that the solid has elastic resistance
to shearing stress while a liquid does not.'' Our work strongly supports
this assertion but points out that the contrast in behaviour between
fluids and crystals shows its greatest effect when one compares the
limiting zero frequency component of the imaginary part of the shear
viscosity. In fluids that value is zero while in crystals the limiting
value is infinity!

\section*{Appendix: Transforming the autocorrelation function}

A trapezoidal version of Filon's quadrature was used to transform
the correlation functions,

\begin{eqnarray}
\int_{t_{0}}^{t_{1}}ds\, f(s)\,\cos(\omega s) & = & \frac{1}{\omega}\boldsymbol{(}f(t_{1})\sin(\omega t_{1})-f(t_{0})\sin(\omega t_{0})\boldsymbol{)}\nonumber \\
 & + & \frac{1}{\omega^{2}}\frac{f(t_{1})-f(t_{0})}{t_{1}-t_{0}}\boldsymbol{(}\cos(\omega t_{1})-\cos(\omega t_{0})\boldsymbol{)},\label{ap2-cos}\end{eqnarray}

\begin{eqnarray}
\int_{t_{0}}^{t_{1}}ds\, f(s)\,\sin(\omega s) & = & \frac{1}{\omega}\boldsymbol{(}f(t_{0})\cos(\omega t_{0})-f(t_{1})\cos(\omega t_{1})\boldsymbol{)}\nonumber \\
 & + & \frac{1}{\omega^{2}}\frac{f(t_{1})-f(t_{0})}{t_{1}-t_{0}}\boldsymbol{(}\sin(\omega t_{1})-\sin(\omega t_{0})\boldsymbol{)}.\label{ap2-sin}\end{eqnarray}
Using these quadrature it is straightforward to prove, given $f(\infty)=0$,
that \begin{equation}
\lim_{\omega\rightarrow\infty}\omega\int_{0}^{\infty}ds\, f(s)\,\cos(\omega s)=0\end{equation}
\begin{equation}
\lim_{\omega\rightarrow\infty}\omega\int_{0}^{\infty}ds\, f(s)\,\sin(\omega s)=f(0)\end{equation}

\begin{acknowledgments}
We thank the Australian Research Council (ARC) for supporting this
work and the National Computational Infrastructure (NCI) for computational
facilities.
\end{acknowledgments}

\end{document}